\documentclass[a4paper,11pt,secnumarabic,amssymb, nobibnotes, prd,footinbib,nofootinbib]{revtex4-1}

\usepackage{geometry}
\usepackage{amsfonts}
\usepackage[pdftex]{graphicx}
\usepackage{multirow}
\usepackage{amsmath}
\usepackage{graphicx}
\usepackage[pdftex]{graphicx}
\usepackage{lipsum}
\usepackage{slashed}
\usepackage{geometry}
\usepackage{multirow}
\usepackage{amsfonts}
\usepackage{placeins}
\usepackage[bottom]{footmisc}
\usepackage[colorlinks=true,linkcolor=red,citecolor=blue]{hyperref}%
\begin{document}

\title{A Curvature Operator for a Regular Tetrahedron Shape in LQG}%
\thanks{This article is based on the work presented in the paper of ref. \cite{28}.}

\author{O.Nemoul}%
\email{omar.nemoul@yahoo.fr}
\address{Laboratoire de physique mathématique et subatomique\\
Mentouri university, Constantine 1, Algeria}

\author{N.Mebarki}%
\email{nnmebarki@yahoo.fr}
\address{Laboratoire de physique mathématique et subatomique\\
Mentouri university, Constantine 1, Algeria}

\date{March 7, 2018}%

\begin{abstract}
An alternative approach introducing a 3 dimensional Ricci scalar
curvature quantum operator given in terms of volume and area as well
as new edge length operators is proposed. An example of
monochromatic 4-valent node intertwiner state (equilateral
tetrahedra) is studied and the scalar curvature measure for a
regular tetrahedron shape is constructed. It is shown that all
regular tetrahedron states are in the negative scalar curvature
regime and for the semi-classical limit the spectrum is very close
to the Euclidean regime.
\end{abstract}

\maketitle
\section{Introduction}

Loop Quantum Gravity (LQG) \cite{1,2} is a background independent
quantum field theory, it has been described as the best way to build
a consistent quantum version of General Relativity (with vanishing
cosmological constant $\Lambda=0$). Canonically, it is based on the
implementation of the Holst action \cite{3} and the Ashtekar-Barbero
variables (The configuration variable is the real $su(2)$ connection
$A^i_a(x)$ and its canonical conjugate is the gravitational electric
field $E^b_j(x)$) with a real Immirzi parameter $\gamma$ \cite{4,5}
by the Dirac quantization procedure \cite{6}. In order to construct
the starting kinematical Hilbert space, one has to use the well
known representation of the holonomy-flux algebra \cite{7}: it is
represented by the space of all cylindrical wave functional through
holonomies defined by the $su(2)$ connection along a system of
smooth oriented paths and flux variables as the smeared electric
field along the dual surface for each path. Due to the
background-independent property of LQG, it was possible to use
Wilson loops \cite{8} which are the natural gauge-invariant holonomy
of the gauge connection as a basis for the gauge invariant Hilbert
space \cite{9}. Another useful basis state of the quantum geometry
known as the Penrose's spin networks is frequently used \cite{10}.
Spin network arises as a generalization of Wilson loops necessary to
deal with mutually intersecting loops "nodes" which is represented
by a space of intertwiners at each node \cite{11}. One can construct
well defined observables such as the area and volume acting on links
and nodes respectively of smooth paths system \cite{12}. The
fuzziness and discreteness property of space \cite{13,14,15} is
predicted. A beautiful interpretation of the intertwiners in terms
of quantum Euclidean polyhedra \cite{16,17} naturally arises. In
this work, we construct a new geometrical information from LQG spin
network based on the polyhedra interpretation of spin network
states, which is the value of the 3d-Ricci scalar curvature and the
edge length as a function of volume and boundary areas operators. A
suggested introduction to the curvature operator in terms of the
length operator and the dihedral angles was provided by using 3d-
Regge calculus \cite{18}. Moreover, there are three proposals for
length operator discussed in refs. \cite{19,20,21}. The main idea of
our work comes from the determination of the volume and the boundary
area of a fixed region in a Riemannian manifold as a function of the
scalar curvature inside that region as well as its parameterization.
One can invert these functions to get the explicit formula of the
scalar curvature in terms of volume and boundary area of a fixed
region. Similar idea can be done using a geodesic polyhedron
shape\footnote{Geodesic polyhedron is the convex region enclosed by
the intersection of geodesic surfaces.  A geodesic surface is a
surface with vanishing extrinsic curvature and the intersection of
two such surfaces is necessarily a geodesic curve.} \cite{28}. By
extending the Euclidean polyhedra interpretation to all non-zero
curvature polyhedra, we can use the new proposed scalar curvature
operator related to a fixed polyhedron measure and try to determine
its spectrum in order to know what kind of space in which the
intertwiner state is represented. This geometrical approach can be
considered as a natural arena for considering LQG including a
cosmological constant. in the case of $\Lambda\neq0$, The $SU(2)$
gauge invariant is still representing the kinematical space of LQG
(since the cosmological constant just appears in the Hamiltonian
constraint). Thus, one can describe the intertwiner state by a
curved chunk of a curved polyhedron and then the main feature of our
proposed curvature operator is to determine in a straightforward
manner which cosmological constant value $\Lambda$ can an
intertwiner state be interpreted as a fixed geodesic polyhedron.
Moreover, a proposal to introduce a non-vanishing cosmological
constant in LQG is to work with the q-deformed
$\mathcal{U}_{q}(su(2))$ rather than the $SU(2)$ itself
\cite{11,22,23,24,25} and the use of curvature tetrahedron was
suggested in \cite{26}. In our approach, an example of a such
monochromatic 4-valent node state was studied in details and its
associated Kapovich-Millson phase space (i.e. the space of all
equilateral Euclidean tetrahedron shapes) was constructed. Moreover,
we will show the absence of a regular Euclidean tetrahedron from the
volume orbit of relevant shapes in that phase space, instead of this
it is possible to find a regular tetrahedron correspondence in the
context of a non-zero constant curvature tetrahedron. It is worth to
mention that the phase space of curved tetrahedron shapes idea has
been initiated in ref. \cite{27}. In our present paper, full
expressions of volume and boundary face area of a regular
tetrahedron in a constant curvature space (in terms of the scalar
curvature and the edge length \cite{28}) are explicitly derived than
inverted to get the exact form of the 3d- Ricci scalar curvature and
the edge length. At the quantum level, we obtain two well defined
operators acting on the monochromatic 4-valent nodes state. Their
spectra show that all quantum atoms of space can be represented by
chunks of regular hyperbolic tetrahedron of a negative curvature
$R\sim -{(8\pi Gh\gamma)}^{-1}$. It also produces the Euclidean
regular tetrahedron $R\sim0$ in the semi-classical limit $j\gg1$
($j$ is links color). In what follows, we will work in a unit where
$8\pi Gh\gamma =1$. The paper is organized as follows : In section
\ref{Sec2}, we give a motivation for a new scalar curvature measure.
In section \ref{Sec3}, a strategy of defining new curvature operator
in LQG is presented. In section \ref{Sec4}, a 3d- Ricci scalar
curvature and edge length operators are constructed for a regular
tetrahedron state. Finally in section \ref{Sec5}, we draw our
conclusions.

\section{ Motivation for a new scalar curvature measure}\label{Sec2}

\begin{figure}[ht]
\begin{center}
\includegraphics[width=2.1in]{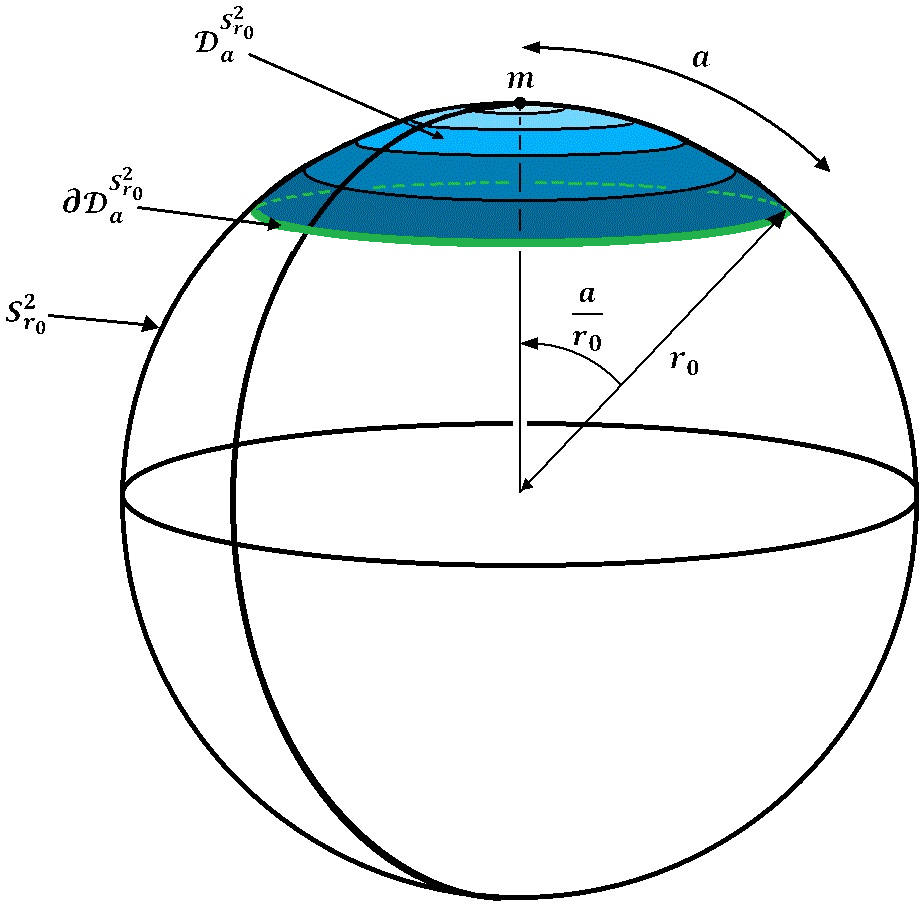}
\end{center}
\caption{The geodesic disc $\mathcal{D}^{S^2_{r_0}}_a$ (blue) and
its boundary circle $\partial{\mathcal{D}}^{S^2_{r_0}}_a$ (green) in
the 2-sphere $S^2_{r_0}$.\label{fig1}}
\end{figure}

General relativity (GR) is a dynamical theory of spacetime within
the framework of the general covariance. Accordingly, it should be
described by geometrical observables with respect to other dynamical
fields\footnote{Due to the diffeomorphism invariance property of GR.
For details see \cite{1}.}. It implies that any invariant measure of
GR can be written as a function of geometrical quantities. For
instance, the 3d- Ricci scalar curvature in some point of the
hypersurface $\Sigma_{t}$ embedded in a smooth Riemannian manifold
$M$ is technically determined by the measure of volume and boundary
area of a neighborhood region around this point. Doing it separately
does not give enough geometrical informations of the space. Rather,
it is mandatory to do this at the same time in order to get the
complete information. To be more explicit, let us consider the
simplest case of the 2-sphere $S^2_{r(t)}$ of radius $r(t)$ in 2+1
dimension (See Fig.~\ref{fig1}). The spatial metric at a given time
$t$ is:
\begin{equation}{{ds}^2|}_{\Sigma_{t}}={r(t)}^2({d\theta }^2+{{\mathrm{sin}}^{\mathrm{2}} (\theta ) }{d\varphi }^2)\,,\label{eq1}\end{equation}
At\ $t=t_0$ we want to measure the 2d- Ricci scalar curvature
$R_{t_0}$ such that  $r(t_0)=r_0$ This means we have to measure the
radius $r_0$ (because $R_{t_0}=\frac{2}{r^2_0}$ ). To do so, we fix
a region ${\mathcal{D}}^{S^2_{r_0}}_a(m)$  of a geodesic disc with a
radius $a$ centering at a point $m\in S^2_{r_0}$:
\begin{equation}{\mathcal{D}}^{S^2_{r_0}}_a\left(m\right)=\left\{p\in S^2_{r_0}\mathrel{\left|\vphantom{p\in S^2_{r_0} l^{S^2_{r_0}}_{mp}\le a}\right.\kern-\nulldelimiterspace}l^{S^2_{r_0}}_{mp}\le a\right\}\subset S^2_{r_0}\,,\label{eq2}\end{equation}
Where  $l^{S^2_{r_0}}_{mp}$  is the geodesic length of the
$S^2_{r_0}$  space between the points $m$ and $p$. The area
$A(r_0,a)$  of the disc and its boundary curve length $L(r_0,a)$
are:
\begin{equation}A(r_0,a)=2\pi r^2_0(1-cos(\frac{a}{r_0}))\,,\label{eq3}\end{equation}
\begin{equation}L(r_0,a)=2\pi r_0\ sin(\frac{a}{r_0})\,,\label{eq4}\end{equation}
Given the pair $(r_0,a)$, one can determine the area of a disc and
its boundary curve length $(A,L)$. It is easy to invert these two
functions to obtain:
\begin{equation} R_{t_0}(A,L)=\frac{2}{r^2_0}=\frac{2(4\pi
A-L^2)}{A^2}\,,\label{eq5}\end{equation}
\begin{equation}a(A,L)=\frac{A}{\sqrt{4\pi A-L^2}}arctan(\frac{L\sqrt{4\pi A-L^2}}{2\pi A-L^2})\,,\label{eq6}\end{equation}
Thus, The simultaneous measurement of the area and the boundary
curve length of a geodesic disc can allows us to estimate the value
of the 2d- Ricci scalar curvature ($R_{t_0}=\frac{2}{r^2_0}$) and
the disc radius $a$\textbf{.}

In 2+1 dimension and for the 2-sphere case, these two relations give
us another way to measure the main important geometrical quantity
which is the value of the 2d- Ricci scalar curvature $R_{t_0}(A,L)$
as a function of the area measure and its boundary curve length of a
disc. Remarkably, this technique does not depend on the choice of
the region; one can choose any shape of a region and get the same
2d- scalar curvature. But how can we generalize this technique for
arbitrary 3-dimensional topological spaces?. To get such a
generalization, we try to find a relationship between the 3d- Ricci
scalar curvature with the measurement of volume and boundary area of
an arbitrary region. It was done by using small geodesic ball
\cite{29}, and for any arbitrary regular tetrahedron in a constant
curvature spaces \cite{28}. The curvature can be determined by
inverting the resulting functions in all cases.
 \FloatBarrier
 \FloatBarrier

\section{Strategy for defining a new curvature operator in LQG}\label{Sec3}
In loop quantum gravity ($\Lambda=0$ or $\Lambda\neq0$)\footnote{We
consider the same $SU(2)$ gauge invariant kinematical Hilbert space
for both $\Lambda=0$ and $\Lambda\neq0$ cases.}, the $SU(2)$
invariant Hilbert space at each F-valent node is the intertwiner
space ${\mathcal{H}}_F\equiv inv(V^{(j_1)}\bigotimes \dots
\bigotimes V^{(j_F)})$ which is the quantization of the
Kapovich-Millson phase space ${\mathcal{S}}_F$ i.e. the space of all
Euclidean polyhedron shapes with fixed F-areas norms ${\{A_f\sim
j_f\}}_{f=\overline{1,F}}$ . This correspondence allows us to
interpret each atom of space on a node (volume eigenstate) as
quantum Euclidean polyhedra states. It offers infinite possible
Euclidean polyhedra shapes for the same intertwiner state. In fact,
after restricting the space of shapes of fixed areas
${\{A_f\}}_{f=\overline{1,F}}$ to a spectrum of volume operator, we
will obtain $(2F-5)$ dimensions hyper-surface of relevant shapes
(since the ${\mathcal{S}}_{\{A_f\}}$ phase space has $2(F-3)$
dimensions). Now, it is legitimate to ask the following questions:

\begin{enumerate}
\item  Is this correspondence unique?
\item  Can Loop Quantum Gravity intertwiners states offer non-zero curvature grains of space?
\item  Can we find other polyhedra shapes possibilities in the non-zero curvature regime? For instance, the absence of the regular Euclidean tetrahedron correspondence with the monochromatic 4-valent node intertwiners, this means that there is no regular tetrahedron belonging to the volume orbits in the space of equilateral tetrahedra shapes; can we find this correspondence in the context of non-zero curvature spaces?
\end{enumerate}

In what follows, we will focus on the two last questions by
considering the generalization of this correspondence. In another
words, we will interpret the intertwiner state by a fixed polyhedron
shape (even if it doesn't belong to the volume orbit of Euclidean
polyhedra shapes) and try to find out what kind of a curved space
one must have in order that this polyhedron grain be nicely
consistent with the area and volume spectra of LQG?. This means, we
seek to find another possibilities of the correspondence in the
context of other non-zero curvature polyhedra shapes. The task now
is to determine new curvature operator related to a fixed polyhedron
shape by using the approach similar to the one mentioned previously
consisting in identifying the volume and areas operators of LQG with
those of the corresponding polyhedron in an arbitrary curved space
and inverting the resulting set of functions to end up to the
classical and quantum formula of scalar curvature related to a fixed
polyhedron. It is worth to mention that the classical consistency of
the 3d- Ricci scalar curvature measure as a function of the volume
and boundary area measures is also well-defined at the quantum level
since the commutativity between their associated geometrical
operators\footnote{In LQG, the volume and area operators are
commute.} is guaranteed in LQG. Unfortunately, we cannot exactly
calculate the volume and boundary face area of a polyhedron in a
general curved space, even if we make a perturbative series
expansion around the Euclidean measure for a small polyhedron as it
was mentioned for the small geodesic ball cases \cite{29}, we don't
have any guidance to estimate the uncertainty of this expansion. The
first problem occurred due the arbitrary degree of freedom of the
considered general curved space. The solution is trivial; one can
just relax the degree of freedom to spaces with a constant scalar
curvature (one degree of freedom)\footnote{The main reason is to
describe LQG with a non-vanishing cosmological constant
$\Lambda=cte$.}. In fact, a spin network state of a fixed graph
(dual to a fixed discretization) induces naturally a discrete
locally valued function of the 3d- Ricci scalar curvature. The
reason is that all quantum geometric operators are not sensitive to
all points inside the quantum atom of space; only nodes and links
represent the quanta of space and its boundary surface respectively.
Thus, each quantum atom of space corresponds to a constant 3d- Ricci
scalar curvature value, i.e. all points inside the quantum atom of
space share the same geometrical property. In the following, we will
make our calculation concerning the volume and boundary area of a
polyhedron in a constant curvature Riemannian manifolds. We remind
that the Riemannian manifolds of a constant curvature can be
classified into the Euclidean $({Euc}^3,R=0)$, spherical $(S^3_{r}\
,R>0)$ and hyperbolic $(H^3_{r},R<0)$ geometries (other spaces that
have a constant curvature are isometric to the one of these three
classes by the Killing-Hopf theorem \cite{30,31}). As a byproduct,
the full expression of volume and boundary face area of a regular
tetrahedron in the 3-sphere $S^3_r$ and the 3-hyperbolic $H^3_r$ has
been derived explicitly in terms of the 3d- Ricci scalar curvature
and the edge length in ref. \cite{28}. In the monochromatic 4-valent
node example, we will be interested to study the possibility of
finding a correspondence with a regular geodesic tetrahedron.
Applying the 3d- Ricci scalar curvature operator related to a
regular tetrahedron region on the intertwiner state for constructing
a space of a constant curvature where one can have the regular
tetrahedron correspondence for any irreducible representation $j$.
\section{Application: A monochromatic 4-valent node state}\label{Sec4}
\subsection{ Quantum equilateral Euclidean tetrahedron}

The corresponding system of a monochromatic 4-valent intertwiner
node is an equilateral Euclidean tetrahedron (tetrahedron with faces
of equal areas, see Fig. \ref{fig2} ) and the main ingredients that
comprise this system can be summarized as follows:
\subsubsection{Intertwiner space ${\mathcal{H}}_{\boldsymbol{4}}$}
 In LQG, the
$SU(2)$ invariant Hilbert space of a monochromatic 4-valent node
($j_1=j_2=j_3=j_4=j$) is the intertwiner space
${\mathcal{H}}_4\equiv inv(V^{(j)}\bigotimes V^{(j)}\bigotimes
V^{(j)}\bigotimes V^{(j)} )$ with a dimension $2j+1$ . There are two
well-defined geometric operators acting on the gauge invariant
intertwiner state $\{|{\otimes }^4_{l=1}j_l,i_K\rangle\} \ (\
l=\overline{1,4},\ K=\overline{0,2j})$:
 \FloatBarrier
\begin{figure}[ht]
\begin{center}
\includegraphics[width=2in]{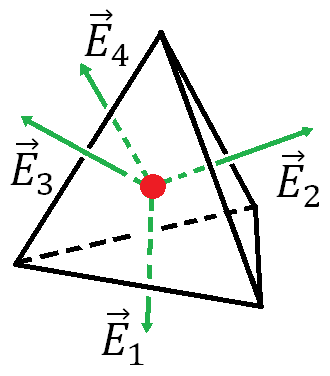}
\end{center}
\caption{Descriptions of the classical geometry of an equilateral
Euclidean tetrahedron.\label{fig2}}
\end{figure}
The area operator acts trivially on the links as:
\begin{equation}{\hat{A}}_f\ |{\otimes }^4_{l=1}j_l,i_K\rangle =\sqrt{{|{\hat{\overrightarrow{E}}}_f|}^2}\ |{\otimes }^4_{l=1}j_l,i_K\rangle \ =\ \sqrt{j(j+1)}\ |{\otimes }^4_{l=1}j_l,i_K\rangle,\,\label{eq7}\end{equation}
The volume operator acts non-trivially on the node \cite{17}:
\begin{equation}
\hat{V} \ |{\otimes }^4_{l=1}j_l,i_K\rangle =\frac{\sqrt{2}}{3}\
\sqrt{|{\hat{\overrightarrow{E}}}_1\cdot
({\hat{\overrightarrow{E}}}_2\times {\hat{\overrightarrow{E}}}_3)|}\
\ |{\otimes }^4_{l=1}j_l,i_K\rangle
\equiv\frac{\sqrt{2}}{3}\sqrt{\hat{Q}}\ |{\otimes
}^4_{l=1}j_l,i_K\rangle\,,\label{eq8}\end{equation}
 We have to diagonalize the
volume matrix element by diagonalizing the matrix $[Q_{K'K}^{(j)}]$
of elements:
\begin{equation}Q_{K'K}^{(j)}\equiv\langle {\otimes }^4_{l=1}j_l,i_{K'}|\
\hat{Q}\ \ |{\otimes }^4_{l=1}j_l,i_K\rangle
\,,\label{eq9}\end{equation}
with
\begin{equation}[Q_{K'K}^{(j)}]=\left(
\begin{array}{ccc}
 \begin{array}{cc}
0 & ia_1 \\
-ia_1 & 0 \end{array}
 & \cdots  & 0 \\
\vdots  & \ddots  & \vdots  \\
0 & \cdots  &  \begin{array}{cc}
0 & ia_{2j+1} \\
-ia_{2j+1} & 0 \end{array}
 \end{array}
\right)\end{equation}

 where
\begin{equation}a_n=\frac{1}{4}\frac{(n^2-{(2j+1)}^2)n^2}{\sqrt{4n^2-1}}\
\ \ \ \ \ \ \ \ n=\overline{1,2j+1}\,,\label{eq10}\end{equation} At
each node, \ the quantum atoms of space is the common
eigenstates\footnote{They are vectors in the intertwiner space
${\mathcal{H}}_{\boldsymbol{4}}$: $|{\otimes
}^4_{l=1}j_l,q_K\rangle=\Sigma^{2j+1}_{K'=1}q_{K}^{K'}|{\otimes
}^4_{l=1}j_l,i_{K'}\rangle$.} $\{|{\otimes
}^4_{l=1}j_l,q_K\rangle\}$ of volume and area operators:
\begin{equation}\hat{V}|{\otimes }^4_{l=1}j_l,q_K\rangle =V_{K}|{\otimes }^4_{l=1}j_l,q_K\rangle \,,\label{eq11}\end{equation}
\begin{equation}\hat{A}|{\otimes }^4_{l=1}j_l,q_K\rangle =\sqrt{j(j+1)}|{\otimes }^4_{l=1}j_l,q_K\rangle\,,\label{eq12}\end{equation}

\subsubsection{The Kapovich-Millson phase space $\boldsymbol{\
}{\mathcal{S}}_{\boldsymbol{4}}$}The space of all Euclidean
equilateral tetrahedron \cite{17} shapes with fixed areas norms
$A_1=A_2=A_3=A_4=A=\sqrt{j(j+1)}$ , satisfying the closure relation:
\begin{equation}{\overrightarrow{A}}_1+{\overrightarrow{A}}_2+{\overrightarrow{A}}_3+{\overrightarrow{A}}_4=\overrightarrow{0}\,,\label{eq13}\end{equation}
The canonical coordinates are:
\begin{equation}p=|{\overrightarrow{A}}_1+{\overrightarrow{A}}_2| \ \ \ \ \ q=\arccos{\frac{({\overrightarrow{A}}_1\times {\overrightarrow{A}}_2)\cdot ({\overrightarrow{A}}_3\times {\overrightarrow{A}}_4)}{|{\overrightarrow{A}}_1\times {\overrightarrow{A}}_2||{\overrightarrow{A}}_3\times {\overrightarrow{A}}_4|}}\,,\label{eq14}\end{equation}
It is obvious that
\begin{equation}0\ \le \ p\ \le \ 2A\ \ \ \ \ \ \ \ \ \ \ \ \ -\frac{\pi }{2}\ \le \ q\ \le \ \frac{\pi }{2}\,,\label{eq15}\end{equation}
All geometrical informations of an Euclidean equilateral tetrahedron
can be constructed from its representation point $(p,q)\in
{\mathcal{S}}_4$ , such as the volume:
\begin{equation}V(A;p,q)=\frac{\sqrt{2}}{3}\sqrt{|{\overrightarrow{A}}_1\cdot ({\overrightarrow{A}}_2\times {\overrightarrow{A}}_3)|}=\frac{1}{3\sqrt{2}}\sqrt{|sin(q)|(\frac{4A^2}{p^2}-1)}\,,\label{eq16}\end{equation}
Notice that the volume function has a maximal value as it is shown
in Fig.~\ref{fig3}. In fact, one has to solve the equations:
\begin{equation}{\frac{\partial V(A;p,q)}{\partial p}|}_{(p_0,q_0)}=0\ \ \ \ \ {\frac{\partial V(A;p,q)}{\partial q}|}_{(p_0,q_0)}=0\,,\label{eq17}\end{equation}
It is easily to check that
\begin{equation}p_0=\frac{2\sqrt{3}}{3}A\ \ \ \ \ \ \ \ \ q_0=\pm \ \frac{\pi }{2}\,,\label{eq18}\end{equation}
where
\begin{equation}V_{max}=V(A;p_0,q_0)=2^{3/2}\ 3^{-7/4}\ A^{3/2}\,,\label{eq19}\end{equation}
which is the expected Euclidean regular tetrahedron.

\subsubsection{The correspondence $\mathcal{H}_{4}\boldsymbol{\leftrightarrow
}\mathcal{S}_{4}$\textbf{}}

\begin{figure}[ht]
\begin{center}
\includegraphics[width=2.3in]{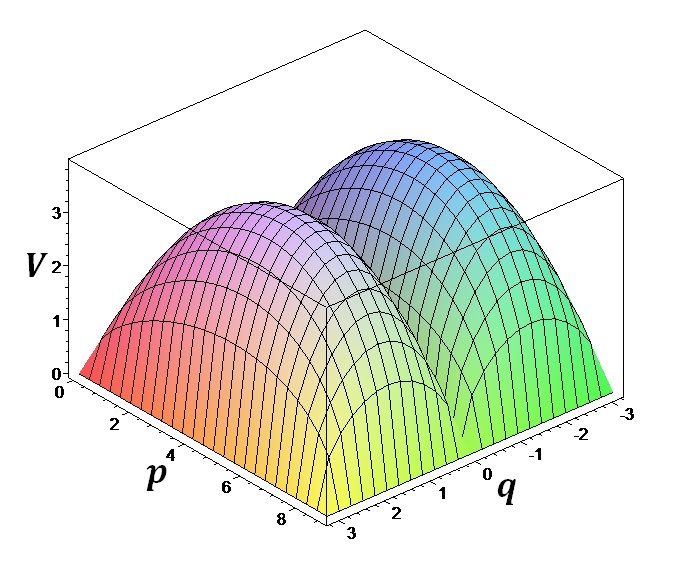}
\end{center}
\caption{The volume function in the the Kapovich-Millson phase space
$\mathcal{S}_{4}$.\label{fig3}}
\begin{center}
\includegraphics[width=2.7in]{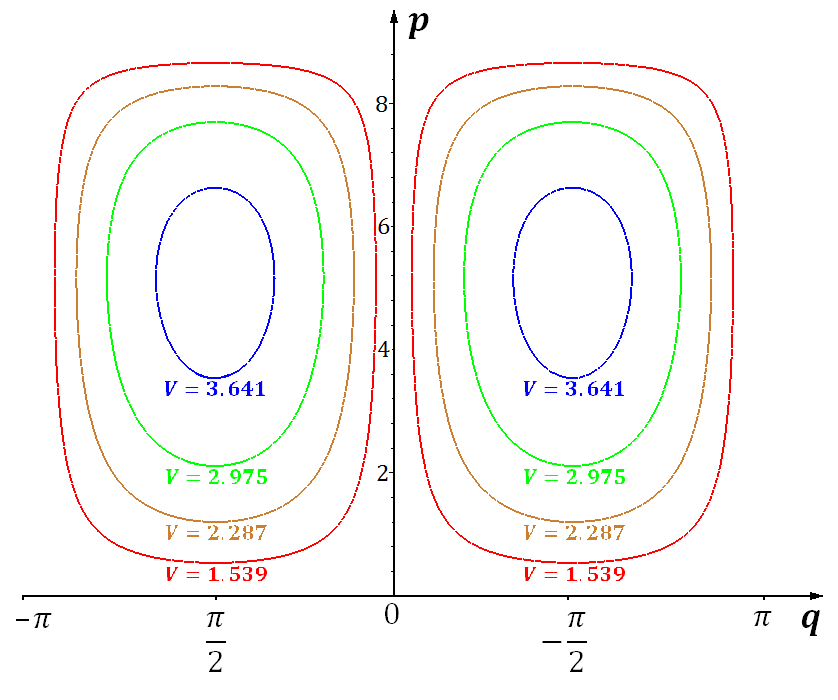}
\end{center}
\caption{The Kapovich-Millson phase space $\mathcal{S}_{4}$. The
colored orbits are quantized levels of the volume operator in the
monochromatic 4-valent eigenstate of $j=4$.\label{fig4}}
\end{figure}

Each volume spectrum (\ref{eq11}) of the intertwiner space
${\mathcal{H}}_4$ corresponds to an orbit in the Kapovich-Millson
phase space ${\mathcal{S}}_4$. These volume orbits are the possible
Euclidean equilateral tetrahedron shapes of the volume eigenstate
with a fixed face area norm $A=\sqrt{j(j+1)}$ (See Fig.~\ref{fig4}).
 \FloatBarrier
 \FloatBarrier

\begin{figure}[ht]
\begin{center}
\includegraphics[width=2.3in]{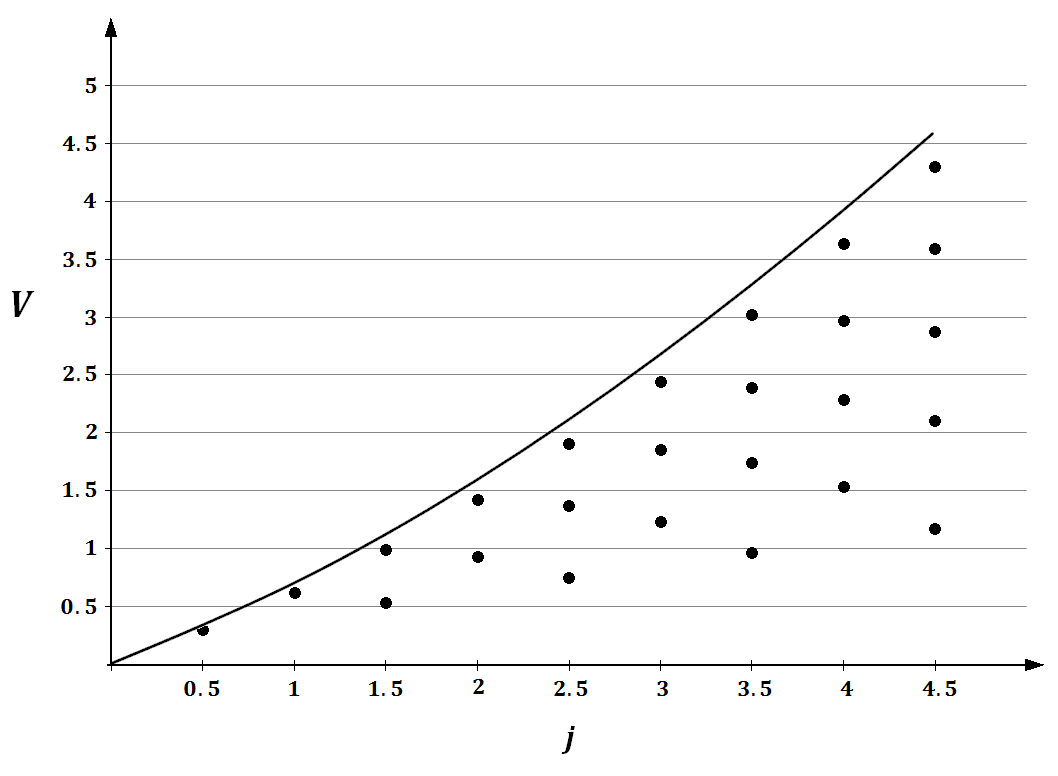}
\end{center}
\caption{Comparison of the regular Euclidean tetrahedron volume
(dark line) with the LQG volume spectra (dots) for the monochromatic
4-valent node state with different links color $j$.\label{fig5}}
\end{figure}

The regular tetrahedron is the only state that has the maximum
volume value. Therefore, the only atom of space state corresponds to
a unique equilateral tetrahedron shape is the one that has a volume
eigenvalue equal to the maximum volume of the phase space
${\mathcal{S}}_F$
\begin{equation}V_{max}=2^{3/2}\ 3^{-7/4}\ {(j(j+1))}^{3/4}\,,\label{eq20}\end{equation}
and it corresponds to the regular tetrahedron.  In LQG, there is no
quantum regular tetrahedron corresponding to a monochromatic
4-valent node state, since all quantum volume spectra are below the
volume of a regular tetrahedron with a face area $A=\sqrt{j(j+1)}$
(See Fig.~\ref{fig5}). The existence of a such regular tetrahedron
solution is guaranteed by the correspondence of the 4-valent node
intertwiner space ${\mathcal{H}}_{\boldsymbol{4}}$\textbf{ }with a
new generalized Kapovich-Millson phase space $\
{\mathcal{S}}_{\{\boldsymbol{4},\boldsymbol{R}\}}$ of equilateral
tetrahedra shapes in constant curvature space $R$ \cite{27}
\begin{equation}{\mathcal{H}}_4\leftrightarrow {\mathcal{S}}_{\{4,R\}}\,,\label{eq21}\end{equation}

\subsection{Ricci scalar curvature and edge length operators for regular
tetrahedron state}

\begin{figure}[h]
\begin{center}
\includegraphics[width=5.7in]{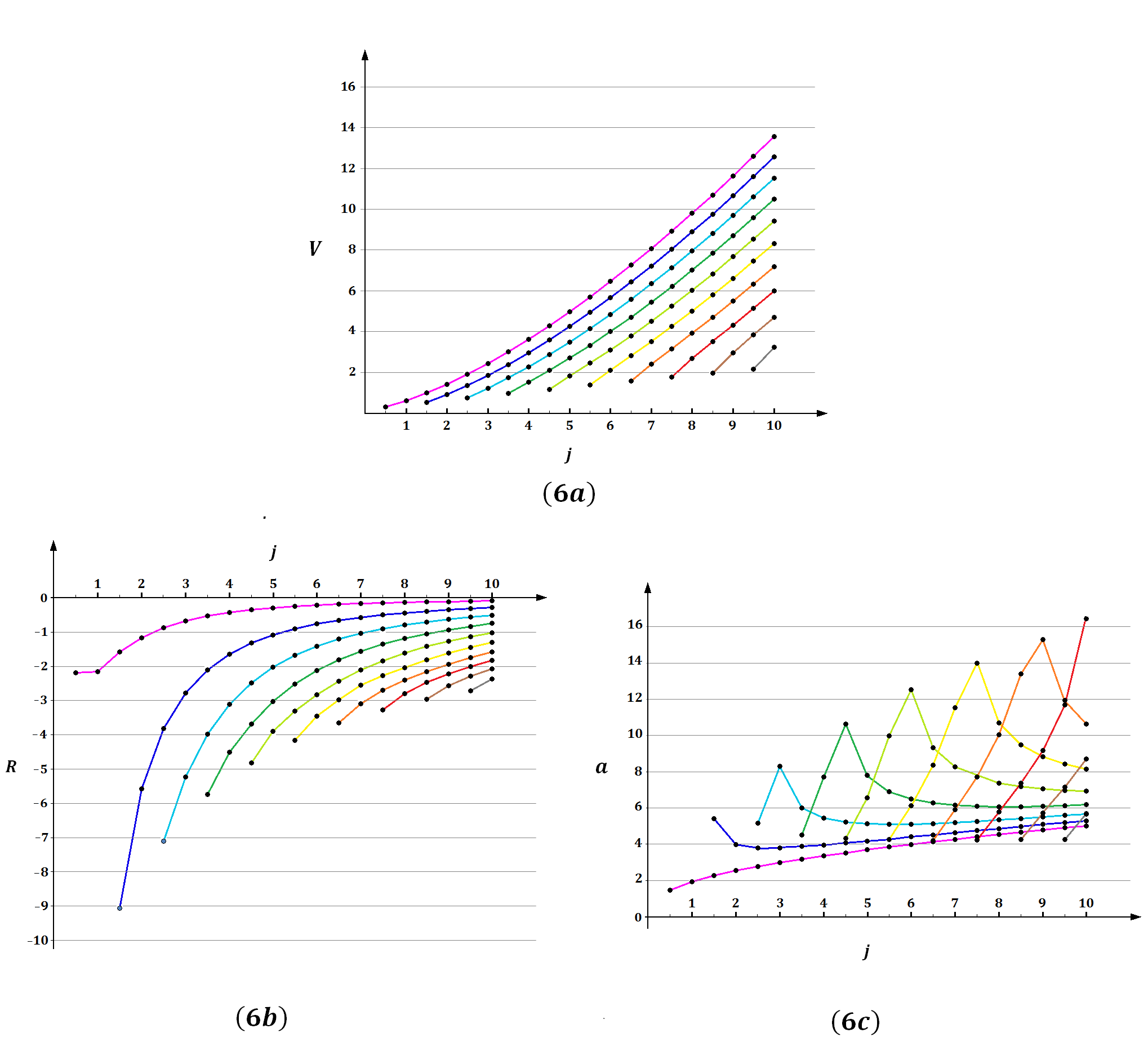}
\end{center}
\caption{Colored lines of different spectra levels for volume (6a),
scalar curvature (6b) and edge length (6c) of a monochromatic
4-valent intertwiner.} \label{fig6}
\end{figure}

Now, let us look for the 3d- Ricci scalar curvature value in which
one can represent the monochromatic 4-valent quanta of space as a
regular tetrahedron in a constant curvature space, in other words we
seek to study the possibility of the regular tetrahedron state
existence in a new equivalent Kapovich-Millson phase space $
{\mathcal{S}}_{\left\{\boldsymbol{4},\boldsymbol{R}\right\}}$ in the
context of constant curvature spaces. In reference \cite{28}, the
volume and the boundary face area of a regular spherical and
hyperbolic tetrahedron given as explicit functions of the edge
length $a$ and the radius $r=\sqrt{\frac{6}{|R|}}$  are shown to
have the following expressions:\footnote{Notice that the geodesic
surfaces of the $S^3_r$ and $H^3_r$ are portions of the great
2-dimensional spheres $S^2_r$ and hyperbolic $H^2_r$ respectively.
Indeed, the area expression (\ref{eq23}) of a regular triangle is a
combination of the area formula given by the dihedral angle $\Theta$
and the cosine rule $\cos(\Theta)=\frac{\cos(\frac{a}{\epsilon
r})}{\cos(\frac{a}{\epsilon r})+1}$ in the context of spherical and
hyperbolic trigonometry.}
\begin{equation}A^{\mathrm{\Sigma }}\left(r,a\right)={\epsilon}^2r^2\left(3\arccos\left(\frac{\cos(\frac{a}{\epsilon r})}{\cos(\frac{a}{\epsilon r})+1}\right)-\pi\right)\,,\label{eq22}\end{equation}
\begin{equation}V^{\mathrm{\Sigma }}(r,a)=12{\epsilon }^3\ r^3\int^{tan(\frac{a}{2\epsilon r})}_0{dt\ \frac{\ t\ arctan(t)}{(3-t^2)\sqrt{2-t^2}}}\,,\label{eq23}\end{equation}
where
\begin{equation}\epsilon =\left\{ \begin{array}{c}
1\ \ \ \ \ \ \ \ \ \ \mathit{\Sigma}=S^3_r \\
i\ \ \ \ \ \ \ \ \ \ \mathit{\Sigma}=H^3_r \end{array} \right.
\,,\label{eq24}\end{equation} \\The Euclidean case is well-defined
in the limit $r\to \infty$ . A direct application of the resulted
formulas (\ref{eq22},\ref{eq23}) in LQG is to find a 3d- scalar
curvature of the quantum atom of space such that the monochromatic
4-valent node has an interpretation of a regular tetrahedron in a
constant curvature space. For each area and volume spectra of the
operators (\ref{eq11},\ref{eq12}), inverting analytically these
systems of functions is not so simple instead, we can deal with it
numerically and construct the 3d- Ricci scalar curvature and the
edge length spectra (See Figs.~\ref{fig6}). In Figs. 6a, 6b and 6c,
each curve with the same color corresponds to volume, scalar
curvature and edge length spectra of the same states
 \FloatBarrier
 \FloatBarrier
 From the the above figures~\ref{fig6} , it is worth to
shed light on the main following conclusions:

\begin{enumerate}
\item The existence of a regular tetrahedron consistent with LQG data
(volume and area spectra) is guaranteed in the negative curvature
regime, and then one can represent the monochromatic 4-valent state
by a regular hyperbolic tetrahedron.

\item In general speaking, the 4-valent monochromatic state that has a
biggest volume represented by a regular tetrahedron in negative
constant curvature space is the closest to the Euclidean space with
the smallest edge length and vice versa.

\item The lowest level value of the edges length (violet curve in Fig. 6c) are
approximately the edges length of the Euclidean regular tetrahedron
with a face area $A=\sqrt{j(j+1)}$
\begin{equation}a_{min}\approx {(\frac{4\ A}{\sqrt{3}})}^{\frac{1}{2}}={(\frac{4\ \sqrt{j(j+1)}}{\sqrt{3}})}^{\frac{1}{2}}\,,\label{eq25}\end{equation}

\item For a generic spin value $j\sim 1$, we find that the regular
tetrahedron solutions of negative scalar curvature spectra are in
the range:
\begin{equation}R\sim -{(8\pi Gh\gamma)}^{-1}\sim -{10}^{70}/\gamma\ m^{-2}\,,\label{eq26}\end{equation}

\item In the semi-classical limit  $j\gg 1$, the monochromatic 4-valent
will be more closer to be identified with the Euclidean regular
tetrahedron, because all scalar curvature spectra vanish as well as
the edge length spectra tend asymptotically to the edge length of a
regular Euclidean tetrahedron given in (\ref{eq25}) (See Figs. 6b,
6c). Accordingly, we are able to have a good approximation of the
volume and boundary face area functions (\ref{eq22},\ref{eq23})
around the zero constant curvature in the case of  $j\gg 1$ . In
fact, by expanding these two functions (\ref{eq22},\ref{eq23}) with
respect to the variable  $\frac{a}{r}$ , we obtain:
\begin{equation}A^{\mathrm{\Sigma }}(r,a)=\frac{\sqrt{3}}{4}a^2[1+\frac{1}{8}{({{\frac{a}{\epsilon r}}})}^2+\mathcal{O}({({{\frac{a}{\epsilon r}}})}^4)]\,,\label{eq27}\end{equation}
\begin{equation}V^{\mathrm{\Sigma }}(r,a)=\frac{\sqrt{2}}{12}a^3[1+\frac{23}{80}{({{\frac{a}{\epsilon r}}})}^2+\mathcal{O}({({{\frac{a}{\epsilon r}}})}^4)]\,,\label{eq28}\end{equation}
\end{enumerate}

As we have previously said, the analytic inversion of the two
functions (\ref{eq22},\ref{eq23}) is not analytically possible,
instead of doing the exact inversion with respect to the exact
variables $(r,a)$, we will use the good approximation functions
(\ref{eq27},\ref{eq28}) with respect to the approximate variables
$(\tilde{r},\tilde{a})$ and write:
\begin{equation}A^{\mathrm{\Sigma }}(\tilde{r},\tilde{a})=\frac{\sqrt{3}}{4}{\tilde{a}}^2(1+\frac{1}{8}\tilde{x})\,,\label{eq29}\end{equation}
\begin{equation}V^{\mathrm{\Sigma }}(\tilde{r},\tilde{a})=\frac{\sqrt{2}}{12}{\tilde{a}}^3(1+\frac{23}{80}\tilde{x})\,,\label{eq30}\end{equation}
where
\begin{equation}\tilde{x}={(\frac{\tilde{a}}{\epsilon \tilde{r}})}^2=\frac{\tilde{R}\ {\tilde{a}}^2}{6}\,,\label{eq31}\end{equation}
Inverting the two functions (\ref{eq29},\ref{eq30}) for the two
variables $\tilde{R}$ and $\tilde{a}$, we obtain approximated
formulas of the scalar curvature as well as the edge length:
\begin{equation}\tilde{R}(A,V)=3\frac{\sqrt{3}}{2A}\tilde{x}(1+\frac{1}{8}\tilde{x})\,,\label{eq32}\end{equation}
\begin{equation}\tilde{a}(A,V)={(\frac{4\sqrt{3}}{3}\frac{A}{1+\frac{1}{8}\tilde{x}})}^{1/2}\,,\label{eq33}\end{equation}
where
\begin{equation}\tilde{x}(A,V)=\frac{4\sqrt{3}\ A}{F(A,V)}-8\,,\label{eq34}\end{equation}
\begin{equation}F(A,V)={[\frac{1}{78}G(A,V)+\frac{23\sqrt{3}A}{G(A,V)}]}^2\,,\label{eq35}\end{equation}
and
\begin{equation}G(A,V)=(-205335\sqrt{3}V+117\sqrt{-1265368\sqrt{3} A^3+9240075V^2})^{1/3}\,,\label{eq36}\end{equation}

 Now, one has to quantize the 3d-
Ricci scalar curvature and edge length functions given in
(\ref{eq32},\ref{eq33}) by quantizing the area and volume operators
to obtain quantum operators that act on the state of monochromatic
4-valent node quantum atom of space (the volume eigenstate):
\begin{equation}\tilde{R}(A,V)\rightarrow\hat{\tilde{R}}(\hat{A},\hat{V})\,,\label{eq37}\end{equation}
\begin{equation}\tilde{a}(A,V)\rightarrow\hat{\tilde{a}}(\hat{A},\hat{V})\,,\label{eq38}\end{equation}
As the color $j$ increases, the accuracy of these two operators
(\ref{eq37},\ref{eq38}) will be very high and their behavior spectra
for $j\to \infty $ in the semi-classical limit is well known and it
gives the Euclidean solution (See table \ref{tab})
\begin{equation}{\hat{\tilde{R}}(\hat{A},\hat{V}) \mathop{|{\otimes }^4_{l=1}j_l,q_K\rangle }_{j\to \infty \mathrm{\ }}\ }=\tilde{R}_{K}(\sqrt{j(j+1)},V_{K})\ \mathop{|{\otimes }^4_{l=1}j_l,q_K\rangle }_{j\to \infty \mathrm{\ }}\approx 0\,,\label{eq39}\end{equation}
\begin{equation}{\hat{\tilde{a}}(\hat{A},\hat{V}) \mathop{|{\otimes }^4_{l=1}j_l,q_K\rangle }_{j\to \infty \mathrm{\ }}\ }=\tilde{a}_{K}(\sqrt{j(j+1)},V_{K})\ \mathop{|{\otimes }^4_{l=1}j_l,q_K\rangle }_{j\to \infty \mathrm{\ }}\approx {(\frac{4\ j}{\sqrt{3}})}^{\frac{1}{2}}\mathop{|{\otimes }^4_{l=1}j_l,q_K\rangle }_{j\to \infty \mathrm{\ }}\,,\label{eq40}\end{equation}

\section{Conclusion}\label{Sec5}

We have found a new approach of measuring the 3d- Ricci scalar
curvature value by measuring the volume of a region and its boundary
area. We have applied this technique in LQG by generalizing the
interpretation of the intertwiner state to all constant curvature
spaces. In the context of a non-vanishing cosmological constant, the
main feature of our proposed curvature operator is to determine in a
straightforward manner which cosmological constant value can an
intertwiner state be interpreted as a geodesic polyhedron. As a
byproduct, we have studied the possibility of finding the regular
tetrahedron correspondence with the monochromatic 4-valent node in
other constant curvature spaces. It is shown that all regular
tetrahedron states are in the negative scalar curvature regime; for
$j\gg1$ the scalar curvature spectrum will be very close to the
Euclidean regime. We conclude that the simultaneous
measure\footnote{At the quantum level, the commutativity is
well-defined between volume and area operators.} of the volume and
the boundary area of the monochromatic 4-valent node state allow us
to estimate the appropriate case of a constant curvature space in
which this state can be interpreted as a regular tetrahedron.
 \FloatBarrier
\begin{table*}[h]
\caption{\label{tab}Comparison of the approximated spectra of the
two operators ($\hat{\widetilde{R}}$,$\hat{\widetilde{a}}$)
associated to a regular tetrahedron with their exact value ($R$,$a$)
for the highest volume level (violet curve in Fig. 6a) of the
monochromatic 4-valent node state for $j=1,2,3,\ldots,10$. }
\begin{ruledtabular}
\begin{tabular}{@{}ccccccccc @{}}  $j$ & $A$ & $V_{max}$ &
$R$ & $\widetilde{R}$ & $\delta{R}\%$ & $a$ & $\widetilde{a}$ &
$\delta{a}\%$ \\ \colrule $1$ & $1.414$ & $0.620$ & $-2.146$ &
$-1.418$ & $34\%$ & $1.954$ & $1.914$ & $2.07\%$ \\  $2$ & $2.449$ &
$1.425$ & $-1.156$ & $-0.782$ & $32\%$ & $2.557$ & $2.511$ & $1.82\%$ \\
$3$ & $3.464$ & $2.444$ & $-0.663$ & $-0.478$ & $28\%$ & $2.998$ &
$2.960$ & $1.25\%$ \\  $4$ & $4.472$ & $3.641$
& $-0,422$ & $-0.320$ & $24\%$ & $3.369$ & $3.340$ & $0.87\%$ \\
$5$ & $5.477$ & $4.990$ & $-0.291$ & $-0.229$ & $21\%$ & $3.700$ &
$3.677$ & $0.63\%$
\\  $6$ & $6.481$ & $6.476$ & $-0.212$ & $-0.172$ & $19\%$
 & $4.003$ & $3.983$ & $0.48\%$ \\  $7$ & $7.483$ & $8.086$ &
$-0.161$ & $-0.134$ & $17\%$ & $4.283$ & $4.267$ & $0.37\%$ \\ $8$ &
$8.485$ & $9.812$ & $-0.127$ & $-0.107$ & $15\%$ & $4.545$ & $4.532$
& $0.30\%$ \\  $9$ & $9.487$ & $11.646$ & $-0.102$ & $-0.088$ &
$14\%$ & $4.793$ & $4.782$ & $0.24\%$ \\  $10$ & $10.488$ & $13.583$
& $-0.084$ & $-0.073$ & $13\%$ & $5.029$ & $5.019$ & $0.20\%$ \\
\end{tabular}
\end{ruledtabular}
\end{table*}

 \FloatBarrier


\begin{thebibliography}{0}
\bibitem{1} C. Rovelli, Quantum gravity, Cambridge university press (2004), \href{http://www.cpt.univ-mrs.fr/~rovelli/book.pdf}{cpt.univ-mrs}.
\bibitem{2} T. Thiemann, Introduction to modern canonical quantum general
relativity, Cambridge University Press (2007),
\href{https://arxiv.org/abs/gr-qc/0110034}{arXiv:gr-qc/0110034}.

\bibitem{3} Holst, Sören, Barbero's Hamiltonian derived from a generalized
Hilbert-Palatini action, Phys. Rev. D. 53 (10): 5966–5969 (1996),
\href{https://arxiv.org/abs/gr-qc/9511026}{arXiv:gr-qc/9511026}.
\bibitem{4} A. Ashtekar, New variables for classical and quantum Gravity, Phys. Rev. Lett. 57, 2244-2247
(1986).
\bibitem{5} J. F. Barbero, Real Ashtekar variables for Lorentzian
signature space times, Phys. Rev. D 51, 5507-5510 (1995),
\href{https://arxiv.org/abs/gr-qc/9410014}{arXiv:gr-qc/9410014}.
\bibitem{6} P. Dirac, Lectures on Quantum Mechanics, (Belfer Graduate School of Science, Yeshiva University
Press,New York 1964).
\bibitem{7} J. Lewandowski, A. Okolow, H. Sahlmann, T. Thiemann, Uniqueness of diffeomorphism invariant states on holonomy—flux algebras, Commun. Math. Phys. 267: 703-733, (2006), \href{https://arxiv.org/abs/gr-qc/0504147}{arXiv:gr-qc/0504147}.
\bibitem{8} Wilson. K, Confinement of quarks, Phys. Rev. D. 10 (8): 2445 (1974).
\bibitem{9} C. Rovelli, L. Smolin, Knot Theory and Quantum Gravity, Phys. Rev. Lett, 61 (10): 1155–1958 (1988).
\bibitem{10} R. Penrose (1971a), Angular momentum: an approach to combinatorial spacetime, in T. Bastin (ed.). R. Penrose (1971b), Applications of negative dimensional
tensors, in D. J. A. Welsh (ed.).
\bibitem{11} C. Rovelli, L. Smolin, Spin networks and quantum gravity, Phys. Rev. D. 52 (10):
5743–5759 (1995),
\href{https://arxiv.org/abs/gr-qc/9505006}{arXiv:gr-qc/9505006}.
\bibitem{12} P. Doná, S. Speziale, Introductory lectures to loop quantum
gravity, lectures given at the 3eme Ecole de Physique Theorique de
Jijel, Algeria (2009),
\href{https://arxiv.org/abs/1007.0402v2}{arXiv:gr-qc/1007.0402v2}.
\bibitem{13} C. Rovelli, L. Smolin, Discreteness of area and volume in quantum gravity, Nucl. Phys. B 442: 593-622 (1995),
\href{https://arxiv.org/abs/gr-qc/9411005}{arXiv:gr-qc/9411005}.
\bibitem{14} A. Ashtekar, J. Lewandowski. Quantum theory of geometry. I: Area operators, Class. Quant. Grav. 14: A55-A82 (1997), \href{https://arxiv.org/abs/gr-qc/9602046}{arXiv:gr-qc/9602046}.
\bibitem{15} A. Ashtekar, J. Lewandowski. Quantum theory of geometry. II: Volume operators. Adv. Theor. Math. Phys. 1: 388-429 (1998), \href{https://arxiv.org/abs/gr-qc/9711031}{arXiv:gr-qc/9711031}.
\bibitem{16} E. Bianchi, P. Dona and S. Speziale, Polyhedra in loop quantum gravity, Phys. Rev. D 83, 044035 (2011),
\href{https://arxiv.org/abs/1009.3402v2}{arXiv:gr-qc/1009.3402v2}.
\bibitem{17} E. Bianchi, Hal M. Haggard, Bohr-sommerfeld quantization of
space, Phys. Rev. D, 86(12): 124010 (2012),
\href{https://arxiv.org/abs/1208.2228}{arXiv:gr-qc/1208.2228}.
\bibitem{18} E. Alesci, M. Assanioussi, J. Lewandowski, A curvature operator for
LQG, Phys. Rev. D 89, 124017 (2014),
\href{https://arxiv.org/abs/1403.3190v4}{arXiv:gr-qc/1403.3190v4}.
\bibitem{19} T. Thiemann, A Length operator for canonical quantum gravity, J. Math. Phys. 39, 3372
(1998),
\href{https://arxiv.org/abs/gr-qc/9606092}{arXiv:gr-qc/9606092}.
\bibitem{20} E. Bianchi, The Length operator in Loop Quantum Gravity, Nucl. Phys. B 807, 591
(2009),
\href{https://arxiv.org/abs/0806.4710}{arXiv:gr-qc/0806.4710}.
\bibitem{21} Y. Ma, C. Soo, J. Yang, New length operator for loop quantum gravity, Phys. Rev. D 81,
124026 (2010),
\href{https://arxiv.org/abs/1004.1063}{arXiv:gr-qc/1004.1063}.
\bibitem{22} V. G. Turaev, O. Y. Viro, State sum invariants of 3 manifolds and quantum 6j symbols," Topology 31 865 (1992).
\bibitem{23} M. Dupuis, F. Girelli, Observables in Loop Quantum Gravity with a cosmological
constant, Phys. Rev. D 90, 104037 (2014),
\href{https://arxiv.org/abs/1311.6841}{arXiv:gr-qc/1311.6841}.
\bibitem{24} Ma. Dupuis, F. Girelli, Quantum hyperbolic geometry in loop quantum gravity with cosmological
constant, Phys. Rev. D87: 121502 (2013),
\href{https://arxiv.org/abs/1307.5461}{arXiv:gr-qc/1307.5461}.
\bibitem{25} V. Bonzom, M. Dupuis, F. Girelli, E. R. Livine, Deformed phase space
for 3d loop gravity and hyperbolic discrete geometries, (2014),
\href{https://arxiv.org/abs/1402.2323}{arXiv:gr-qc/1402.2323}.
\bibitem{26} Y. Taylor and C. Woodward, 6j symbols for Uq (sl2) and non-Euclidean tetrahedra, Sel. Math. New. Ser. 11,
539 (2005),
\href{https://arxiv.org/abs/math/0305113}{arXiv:math.QA/0305113}.
\bibitem{27} H. M. Haggard, Mu. Han, A. Riello, Encoding Curved Tetrahedra in Face Holonomies: a Phase Space of
Shapes from Group-Valued Moment Maps, Annales Henri Poincar\'e 17
no.8, 2001-2048 (2016),
\href{https://arxiv.org/abs/1506.03053}{arXiv:gr-qc/1506.03053}.
\bibitem{28} O. Nemoul, N. Mebarki, Volume and Boundary Face Area of a Regular Tetrahedron in a Constant Curvature Space,(2018), \href{https://arxiv.org/abs/1803.10809}{arXiv:gen-ph/1803.10809}.
\bibitem{29} A. Gray, The volume of a small geodesic ball of a Riemannian
manifold, Michigan Math. J. 20 (1974).
\bibitem{30} H. Hopf, Zum Clifford-Kleinschen Raumproblem, Mathematische Annalen, Springer
Berlin Heidelberg, 95: 313–339 (1926).
\bibitem{31} W. Killing, Ueber die
Clifford-Klein'schen Raumformen, Mathematische Annalen, Springer
Berlin Heidelberg, 39: 257–278 (1891).


\end{thebibliography}
\end{document}